# Quantum secure direct communication with SU(2) invariant separable polarisation-OAM states

*Sooryansh Asthana\*, Rajni Bala, V. Ravishankar*

*Department of Physics, IIT Delhi, Hauz Khas, New Delhi, 110016, Delhi, India*

*\*sooryansh.asthana@physics.iitd.ac.in*

We propose a quantum secure direct communication protocol using SU (2)- invariant $2 \times N$ separable states, identified as separable equivalents of two-qubit entangled Werner states in [Bharath & Ravishankar, Phys. Rev. A 89, 062110}]. These states can be experimentally realised as incoherent superpositions of separable polarisation-OAM states.

## 1. Introduction

Quantum secure direct communication (QSDC) protocols are implemented by employing photonic qubits [1, 2]. In entanglement-based QSDC protocols, the resource entangled states are prepared through spontaneous-parametric down conversion, whose yield is very low (of the order of 4 pairs per $10^{-6}$ pump photons) [3]. Recently, it has been observed that in noisy-intermediate-scale-quantum computing (NISQ) era, imperfection of devices should be embedded in realistic protocols [4].

Motivated by this, we propose a QSDC protocol by employing separable $2 \times N$ dimensional states. These separable states have been identified as equivalent *states* [5], which can be experimentally realised, thanks to experimental advances in orbital angular momentum (OAM) states (see, for example, [6] and references therein). In fact, we have shown [7] that these states can be prepared by incoherent superpositions of only six pure polarisation-OAM separable SU (2) coherent states to a very good approximation. The polarisation-OAM separable pure SU (2) coherent states have been prepared with 20% efficiency (to be compared with $\approx 10^{-6}$ in SPDC) [8].

## 2. Preliminaries

In this section, we briefly recapitulate the formalism involving equivalent states. For a detailed discussion, one can refer to [5].

### I. SU(2) coherent states (SCS) and the Q-representation

Suppose that $S_1, S_2, S_3$ represent the three generators of (2S+1)- dimensional irreducible representation of SU (2) group. A (2S+1)- dimensional SCS $|\hat{n}(\theta, \phi)\rangle$ may be generated by the action of (SU (2)) group on the state, $|S_3 = S\rangle$ [9],

$$|\hat{n}(\theta, \phi)\rangle \equiv e^{-iS_3\phi} e^{-iS_2\theta} e^{-iS_1\psi}|S\rangle.$$

The set of all SCSs $\{|\hat{n}(\theta, \phi)\rangle\}$ forms an overcomplete set. Being over-complete, SCSs allow any state to be *completely* expressed in terms of its diagonal elements, $F(\hat{n}) = \frac{2S+1}{4\pi}\langle\hat{n}|\rho|\hat{n}\rangle$, termed as *the Q-representation*.

## II. Equivalent states and equivalent observables

Two states belonging to two Hilbert spaces of different dimensions are termed as *equivalent* if they have the same Q-representation. For example, consider two Hilbert spaces $H_{d_1}$ and $H_{d_2}$ of dimensions $d_1$ and $d_2$ respectively ($d_1 < d_2$). If a state $\rho \in H_{d_1}$ has an equivalent state $\rho' \in H_{d_2}$, then for an observable $O \in H_{d_1}$, there exists $O' \in H_{d_2}$ such that, $Tr(\rho O) = Tr(\rho' O')$. The observables $O$ and $O'$ are termed as *equivalent observables*.

## III. Classical simulation of entangled states

The equivalence between lower dimensional entangled states and higher-dimensional separable states is termed as *classical simulation* of entangled states [5]. For example, the $2 \times (2S+1)$ dimensional equivalent of $2 \times 2$ dimensional Werner state,

$$\rho_W^{\{2,2\}}(\alpha) \equiv \frac{1}{4}(1 - \alpha \vec{\sigma}^A \cdot \vec{\sigma}^B); \alpha \in \left[-\frac{1}{3}, 1\right],$$

is given by

$$\rho_W^{\{2,2S+1\}}(\alpha) \equiv \frac{1}{2(2S+1)}(1 - \alpha \vec{\sigma}^A \cdot \vec{S}^B); \alpha \in \left[-\frac{S}{S+1}, 1\right].$$

While the state $\rho_W^{\{2,2\}}(\alpha)$ is entangled in the range $\frac{1}{3} \leq \alpha \leq 1$, its equivalent $\rho_W^{\{2,2S+1\}}(\alpha)$ is entangled in the range $\frac{S}{S+1} \leq \alpha \leq 1$. So, a $2 \times 2$ entangled Werner state has a $2 \times 2S+1$ separable equivalent state, for an appropriately chosen $S$. The eigenstates of $\sigma_3^A$ and $S_3^B$ may be identified with two polarisation states of light and (2S+1) orthonormal Laguerre-Gauss modes respectively.

## IV. Equivalent of CHSH inequality

The $2 \times 2S+1$ equivalent of CHSH inequality is given by,

$$\frac{3}{S+1} |\langle \vec{\sigma}^A \cdot \hat{a}_1 \, \hat{S}^B \cdot (\hat{b}_1 + \hat{b}_2) + \vec{\sigma}^A \cdot \hat{a}_2 \, \hat{S}^B \cdot (\hat{b}_1 - \hat{b}_2) \rangle| \leq 2 \qquad \cdots (1)$$

### 3. QSDC protocol with polarisation--OAM separable equivalent states

In this section, we present the steps of the QSDC protocol with $2 \times N$ separable equivalent states.

1. Alice and Bob share a bipartite polarisation-OAM system in the state $\rho_W^{[2,2S+1]}(\alpha)$. Alice measures one of the observables from $\sigma_1^A$, $\frac{\sigma_1^A + \sigma_3^A}{\sqrt{2}}$, $\sigma_3^A$ on her polarisation qubit randomly with an equal probability. This can be implemented through a tuneable polariser.
2. Bob measures one of the observables from $\hat{S}_3^B$, $\frac{S_1^B \pm S_3^B}{\sqrt{2}}$ on his OAM qudit randomly with an equal probability. This can be implemented by employing spatial light modulators.
3. *Security check:* Alice and Bob check for violation of the equivalent CHSH inequality (1) as a security check against eavesdropping.

4. *Generation of correlated bit string:* If eavesdropping is so ruled out, then, (i) Alice chooses the data of those rounds in which she has measured $\sigma_1^A$, and $\frac{\sigma_1^A+\sigma_3^A}{\sqrt{2}}$. (ii) Bob chooses the data of those rounds in which he has measured $\hat{S}_1^B$ or $\frac{S_1^B+S_3^B}{\sqrt{2}}$ and obtained outcomes $\pm 1$. This data results in a string of correlated bits between Alice and Bob.
5. In order to send the message, Alice performs operations $\mathbb{1}$ (no transformation) or $X \equiv |0\rangle\langle 1| + |1\rangle\langle 0|$ and send the transformed bits to Bob. Bob, upon receiving the bits, measures them and compares with his outcomes obtained in step (4). The transformations $\mathbb{1}$ and $X$ correspond to bits 0 and 1 respectively.

This concludes the protocol.

*Key generation rate:* The sifted key rate of the protocol is given by mutual information between Alice and Bob. The mutual information is given by $H(A) + H(B) - H(A,B) = \frac{1+\alpha}{2}\log(1+\alpha) + \frac{1-\alpha}{2}\log(1-\alpha)$ bits per transmission.

## 4. Conclusion

We have proposed a QSDC protocol by employing only mixed separable polarisation-OAM states. These states can be realised with current technologies. Our protocol bypasses the need for spontaneous parametric down conversion, whose yield is very low (of the order of mHz).